\documentclass[twocolumn]{aastex63}

\usepackage{amsmath}
\usepackage{float}

\newcommand{\dm}[2][=]{$\text{DM} #1 #2$\,pc\,cm$^{-3}$}
\newcommand{\rotm}[2][=]{$\text{RM} #1 #2$\,rad\,cm$^{-2}$}

\shorttitle{Localization of repeating FRBs by CHIME/FRB}
\shortauthors{}
\graphicspath{{./}{figures/}}

\begin{document}

\title{Sub-arcminute localization of 13 repeating fast radio bursts detected by CHIME/FRB}

\correspondingauthor{Daniele Michilli}
\email{danielemichilli@gmail.com}

\author[0000-0002-2551-7554]{Daniele Michilli}
  \affiliation{MIT Kavli Institute for Astrophysics and Space Research, Massachusetts Institute of Technology, 77 Massachusetts Ave, Cambridge, MA 02139, USA}
  \affiliation{Department of Physics, Massachusetts Institute of Technology, 77 Massachusetts Ave, Cambridge, MA 02139, USA}
\author[0000-0002-3615-3514]{Mohit Bhardwaj}
  \affiliation{Department of Physics, McGill University, 3600 rue University, Montr\'eal, QC H3A 2T8, Canada}
  \affiliation{Trottier Space Institute at McGill University, 3550 rue University, Montr\'eal, QC H3A 2A7, Canada}
  \affiliation{Department of Physics, Carnegie Mellon University, 5000 Forbes Avenue, Pittsburgh, 15213, PA, USA}
\author[0000-0002-1800-8233]{Charanjot Brar}
  \affiliation{Department of Physics, McGill University, 3600 rue University, Montr\'eal, QC H3A 2T8, Canada}
  \affiliation{Trottier Space Institute at McGill University, 3550 rue University, Montr\'eal, QC H3A 2A7, Canada}
\author[0000-0002-3382-9558]{B.~M.~Gaensler}
  \affiliation{Dunlap Institute for Astronomy and Astrophysics, 50 St. George Street, University of Toronto, ON M5S 3H4, Canada}
  \affiliation{David A. Dunlap Department of Astronomy and Astrophysics, 50 St. George Street, University of Toronto, ON M5S 3H4, Canada}
\author[0000-0001-9345-0307]{Victoria M.~Kaspi}
  \affiliation{Department of Physics, McGill University, 3600 rue University, Montr\'eal, QC H3A 2T8, Canada}
  \affiliation{Trottier Space Institute at McGill University, 3550 rue University, Montr\'eal, QC H3A 2A7, Canada}
\author[0000-0002-8139-8414]{Aida Kirichenko}
  \affiliation{Instituto de Astronom\'{i}a, Universidad Nacional Aut\'{o}noma de M\'{e}xico, Apdo. Postal 877, Ensenada, Baja California 22800, M\'{e}xico}
  \affiliation{Ioffe Institute, 26 Politekhnicheskaya St., St. Petersburg 194021, Russia}
\author[0000-0002-4279-6946]{Kiyoshi W.~Masui}
  \affiliation{MIT Kavli Institute for Astrophysics and Space Research, Massachusetts Institute of Technology, 77 Massachusetts Ave, Cambridge, MA 02139, USA}
  \affiliation{Department of Physics, Massachusetts Institute of Technology, 77 Massachusetts Ave, Cambridge, MA 02139, USA}
\author[0000-0003-3367-1073]{Chitrang Patel}
  \affiliation{Department of Physics, McGill University, 3600 rue University, Montr\'eal, QC H3A 2T8, Canada}
\author[0000-0003-3154-3676]{Ketan R.~Sand}
  \affiliation{Department of Physics, McGill University, 3600 rue University, Montr\'eal, QC H3A 2T8, Canada}
  \affiliation{Trottier Space Institute at McGill University, 3550 rue University, Montr\'eal, QC H3A 2A7, Canada}
\author[0000-0002-7374-7119]{Paul Scholz}
  \affiliation{Dunlap Institute for Astronomy and Astrophysics, 50 St. George Street, University of Toronto, ON M5S 3H4, Canada}
\author[0000-0002-6823-2073]{Kaitlyn Shin}
  \affiliation{MIT Kavli Institute for Astrophysics and Space Research, Massachusetts Institute of Technology, 77 Massachusetts Ave, Cambridge, MA 02139, USA}
  \affiliation{Department of Physics, Massachusetts Institute of Technology, 77 Massachusetts Ave, Cambridge, MA 02139, USA}
\author[0000-0001-9784-8670]{Ingrid Stairs}
  \affiliation{Department of Physics and Astronomy, University of British Columbia, 6224 Agricultural Road, Vancouver, BC V6T 1Z1 Canada}
\author[0000-0003-2047-5276]{Tomas Cassanelli}
  \affiliation{Department of Electrical Engineering, Universidad de Chile, Av. Tupper 2007, Santiago 8370451, Chile}
\author[0000-0001-6422-8125]{Amanda M.~Cook}
  \affiliation{Dunlap Institute for Astronomy and Astrophysics, 50 St. George Street, University of Toronto, ON M5S 3H4, Canada}
  \affiliation{David A. Dunlap Department of Astronomy and Astrophysics, 50 St. George Street, University of Toronto, ON M5S 3H4, Canada}
\author[0000-0001-7166-6422]{Matt Dobbs}
  \affiliation{Department of Physics, McGill University, 3600 rue University, Montr\'eal, QC H3A 2T8, Canada}
  \affiliation{Trottier Space Institute at McGill University, 3550 rue University, Montr\'eal, QC H3A 2A7, Canada}
\author[0000-0003-4098-5222]{Fengqiu Adam Dong}
  \affiliation{Department of Physics and Astronomy, University of British Columbia, 6224 Agricultural Road, Vancouver, BC V6T 1Z1 Canada}
\author[0000-0001-8384-5049]{Emmanuel Fonseca}
  \affiliation{Department of Physics and Astronomy, West Virginia University, PO Box 6315, Morgantown, WV 26506, USA }
  \affiliation{Center for Gravitational Waves and Cosmology, West Virginia University, Chestnut Ridge Research Building, Morgantown, WV 26505, USA}
\author[0000-0003-2405-2967]{Adaeze Ibik}
  \affiliation{Dunlap Institute for Astronomy and Astrophysics, 50 St. George Street, University of Toronto, ON M5S 3H4, Canada}
  \affiliation{David A. Dunlap Department of Astronomy and Astrophysics, 50 St. George Street, University of Toronto, ON M5S 3H4, Canada}
\author[0000-0003-4810-7803]{Jane Kaczmarek}
  \affiliation{Dominion Radio Astrophysical Observatory, Herzberg Research Centre for Astronomy and Astrophysics, National Research Council Canada, PO Box 248, Penticton, BC V2A 6J9, Canada}
\author[0000-0002-4209-7408]{Calvin Leung}
  \affiliation{MIT Kavli Institute for Astrophysics and Space Research, Massachusetts Institute of Technology, 77 Massachusetts Ave, Cambridge, MA 02139, USA}
  \affiliation{Department of Physics, Massachusetts Institute of Technology, 77 Massachusetts Ave, Cambridge, MA 02139, USA}
\author[0000-0002-8912-0732]{Aaron B.~Pearlman}
  \affiliation{Department of Physics, McGill University, 3600 rue University, Montr\'eal, QC H3A 2T8, Canada}
  \affiliation{Trottier Space Institute at McGill University, 3550 rue University, Montr\'eal, QC H3A 2A7, Canada}
\author[0000-0002-9822-8008]{Emily Petroff}
  \affiliation{Department of Physics, McGill University, 3600 rue University, Montr\'eal, QC H3A 2T8, Canada}
  \affiliation{Trottier Space Institute at McGill University, 3550 rue University, Montr\'eal, QC H3A 2A7, Canada}
\author[0000-0002-4795-697X]{Ziggy Pleunis}
  \affiliation{Dunlap Institute for Astronomy and Astrophysics, 50 St. George Street, University of Toronto, ON M5S 3H4, Canada}
\author[0000-0002-4795-697X]{Masoud Rafiei-Ravandi}
  \affiliation{Department of Physics, McGill University, 3600 rue University, Montr\'eal, QC H3A 2T8, Canada}
  \affiliation{Trottier Space Institute at McGill University, 3550 rue University, Montr\'eal, QC H3A 2A7, Canada}
\author[0000-0001-5504-229X]{Pranav Sanghavi}
  \affiliation{Department of Physics, Yale University, New Haven, CT 06520, USA}
\author[0000-0003-2548-2926]{Shriharsh P.~Tendulkar}
  \affiliation{Department of Astronomy and Astrophysics, Tata Institute of Fundamental Research, Mumbai, 400005, India}
  \affiliation{National Centre for Radio Astrophysics, Post Bag 3, Ganeshkhind, Pune, 411007, India}



\begin{abstract}
We report on improved sky localizations of thirteen repeating fast radio bursts (FRBs) discovered by CHIME/FRB via the use of interferometric techniques on channelized voltages from the telescope. 
These so-called `baseband localizations' improve the localization uncertainty area presented in past studies by more than three orders of magnitude.
The improved localization regions are provided for the full sample of FRBs to enable follow-up studies.
The localization uncertainties, together with limits on the source distances from their dispersion measures (DMs), allow us to identify likely host galaxies for two of the FRB sources.
FRB~20180814A lives in a massive passive red spiral at $z \sim 0.068$ with very little indication of star formation, while FRB~20190303A resides in a merging pair of spiral galaxies at $z \sim 0.064$ undergoing significant star formation.
These galaxies show very different characteristics, further confirming the presence of FRB progenitors in a variety of environments even among the repeating sub-class.
\end{abstract}

\keywords{}


\section{Introduction} \label{sec:intro}
Fast radio bursts \citep[FRBs;][]{lor07,tho13} are a class of fast radio transients visible from distant galaxies.
Despite the detection of FRB-like radio signals from a Galactic magnetar \citep{chi20_galactic_FRB,boc20}, the physical origin of the FRB population is still debated, and many models are still viable \citep{pla18}.
Models of FRB progenitors can be constrained by studying the host galaxies of FRB sources \citep[e.g.][]{bha20,hei20}.
Precise localizations of FRB sources observed to emit multiple bursts \citep{spi16} allow follow-up observations with instruments having a relatively small field of view \citep[e.g.][]{cha17_R1,mar20}.
In this respect, repeating FRB sources located in the local Universe are particularly interesting for sensitive multi-wavelength observations \citep[e.g.][]{sch17}.
Also, FRBs localized to specific host galaxies can be used as cosmological probes because their redshift can be compared to the dispersion induced in their signal by the integrated column density of free electrons along the line-of-sight, quantified by their dispersion measure (DM, \citealp{mcq14}).
\citet{mac20} used 8 FRB sources localized to specific galaxies to confirm the presence of the Universe's so-called `missing baryons' in the intergalactic medium (IGM).
Despite the applications described above, only roughly two dozen FRB sources have been localized to specific host galaxies so far \citep{hei20}.
This is due to the challenges of detecting a large number of FRBs using instruments with high angular resolution.
A new generation of telescopes is currently under development to overcome this limitation \citep[e.g.][]{hal19_dsa2000,leu21}. 

The Canadian Hydrogen Intensity Mapping Experiment (CHIME; \citealp{chi22}) is detecting hundreds of FRBs per year \citep{chi21_catalog} thanks to its dedicated CHIME/FRB backend \citep{chi18_overview}.
A fraction of these FRBs have been observed to repeat \citep[e.g.][]{chi19_r2,chi19_8repeaters,fon20}.
We collected raw channelized `baseband' data for some of these repeating FRB sources.
By using these baseband data, the localization precision can be improved up to $\sim 11\arcsec$ for bright bursts \citep{mic21}, an area $\gtrsim 3$ orders of magnitude smaller than measured at discovery and presented in previous studies.
Host galaxy associations have already been presented for five of the repeating FRBs discovered by CHIME/FRB; these are FRBs 
20180916B \citep{mar20};
20181030A \citep{bha21_r4}; 
20200120E \citep{bha21_m81};
20201124A \citep{nim22}; and
20201124A \citep{mar22}.
FRBs 20181030A and 20200120E have been localized with the same CHIME/FRB baseband method presented here, while FRBs 20180916B and 20201124A associations have been obtained with very long-baseline interferometry (VLBI) using the source position obtained by CHIME/FRB's baseband pipeline as an initial parameter.
Currently, the CHIME/FRB Outriggers project is under development to improve the localization precision of CHIME to $\sim 50$\,mas using three additional radio telescopes \citep{leu21,cas22,men22}. 
Until these outrigger telescopes are available, however, the localization precision obtainable from baseband data remains the state-of-the-art for most FRBs discovered by CHIME. 

In the following, we present the positions of 13 repeating FRBs discovered by CHIME/FRB \citep{chi19_r2,chi19_8repeaters,fon20} refined by using stored baseband data \citep{mic21} and a likely host association for two of them.
The polarization properties of the bursts are presented by \citet{mck22}.
A summary of the observational setup and data analysis is described in \textsection\ref{sec:obsertations_analysis} and the source positions are presented in \textsection\ref{sec:results}, together with proposed galaxy associations for two FRBs.
The implications of these new localizations are discussed in \textsection\ref{sec:discussion}, and conclusions are drawn in \textsection\ref{sec:conclusions}.

\section{Observations and data analysis}
\label{sec:obsertations_analysis}
The hardware and design of CHIME is discussed by \citet{chi22}, the detection of FRBs with CHIME/FRB and the capture of baseband data are described in detail by \citet{chi18_overview}, the pipeline to process such baseband data is presented by \citet{mic21}, and the method used to identify possible hosts is outlined by \citet{bha21_r4}.
A summary of these steps is given here, together with a description of the optical observations used to refine the redshift of galaxies in the FRB fields.

\subsection{FRB detection and baseband data}
CHIME is a radio telescope composed of 4 cylindrical reflectors orientated South-North and instrumented with 1024 dual-polarization antennas that monitor the sky over $400$-$800$\,MHz.
An FPGA-based F-engine digitizes the data and applies a polyphase filter bank that divides the bandwidth into 1024 frequency channels and produces the channelized baseband data used in this study.
Subsequently, a GPU-based X-engine correlates the signal from different antennas and the CHIME/FRB backend searches for FRB-like signals in the total intensity data from 1024 FFT beams covering the telescope's field-of-view.
When a candidate event is detected by this real-time detection pipeline, $\sim 100$\,ms of baseband data recorded by the F-engine are stored around the burst.

\subsection{Baseband localization}
An offline pipeline is run on the baseband data to automatically produce a number of scientific outputs. 
Among these, the initial source localization is refined with interferometric techniques. 
A grid of 5$\times$5 beams is formed around the initial localization and a signal-to-noise ratio (S/N) value is calculated for each of them.
The resulting intensity map of the signal is fitted with a 2D Gaussian function approximating the sensitivity response of CHIME's formed beams.
We used a sample of sources with known positions at different sky positions to assess the localization capability of the baseband processing pipeline and to estimate the impact of unaccounted systematic effects.
With a least-squares fit, we measured a calibration for our localizations $\theta$ and their uncertainties $\sigma$ given by \citep{mic21}
\begin{align} \label{eq:loc_calibration}
    \theta^i_x &\pm \sigma^i_x \rightarrow \left(\theta^i_x + 0.16\arcmin\right) \pm \left(\sqrt{\left(1.1 \sigma^i_x\right)^2 + 0.19\arcmin^2}\right)
    \nonumber \\
    \theta^i_y &\pm \sigma^i_y \rightarrow \left(\theta^i_y + 0.17\arcmin\right) \pm \left(\sqrt{\left(1.1 \sigma^i_y\right)^2 + 0.19\arcmin^2}\right),
\end{align}
where $x$ and $y$ are celestial coordinates centered on CHIME and running in the East-West ($x$) and South-North ($y$) directions.
All the positions presented in this paper have been corrected with this calibration.
For each source, we calculated a weighted mean position based on the localization region of single bursts, accounting for the systematics defined in Eq.~\ref{eq:loc_calibration}.

\begin{figure}
\epsscale{1.2}
\plotone{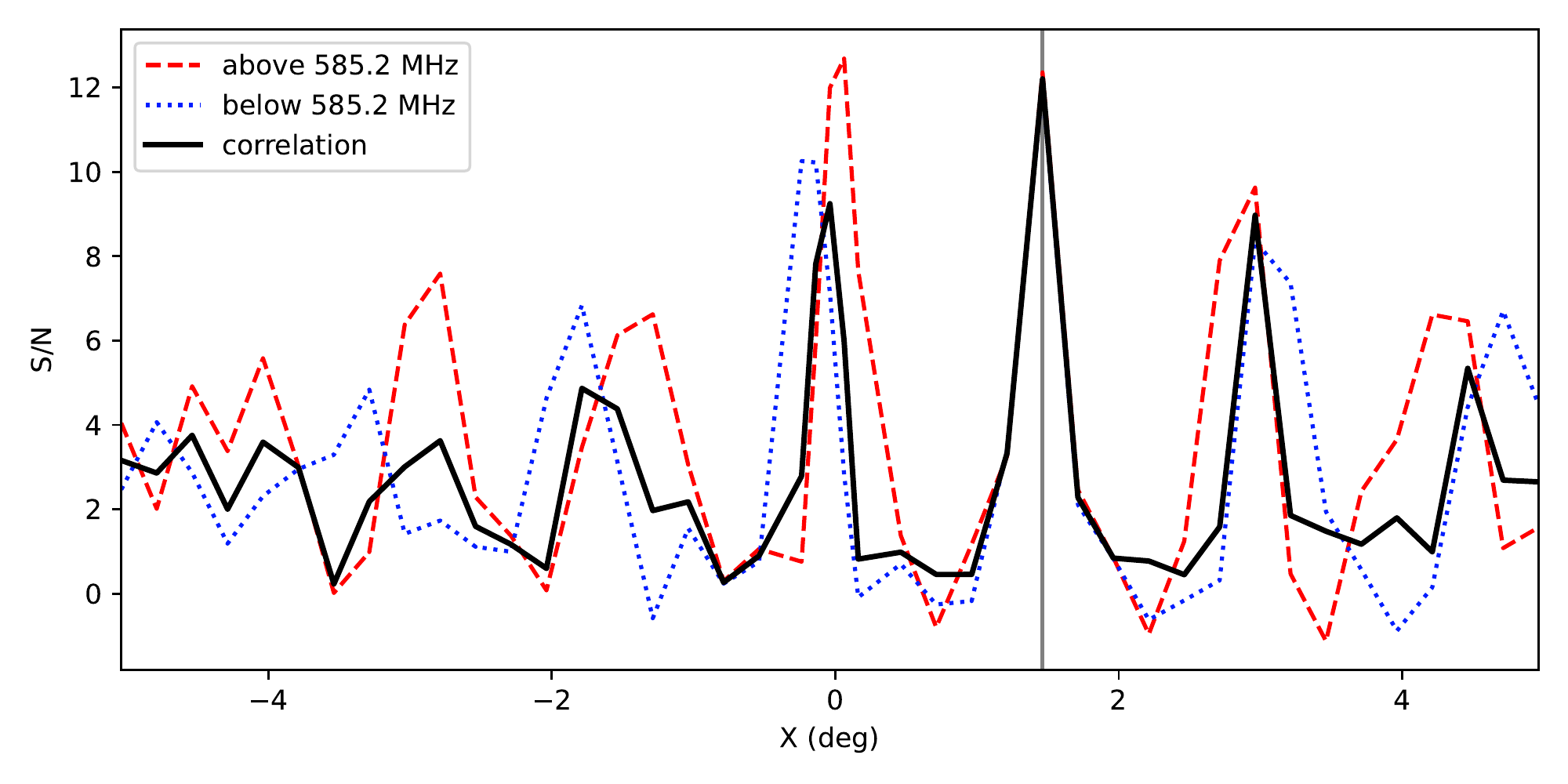}
\caption{\label{fig:sidelobes}
  Signal strength of FRB 20190117A, burst 2, as a function of the local coordinate $x$, centered on CHIME and running from East to West, in degrees of arc. 
  The signal has been divided into two frequency bands around a frequency $\nu_0=585.2$\,MHz, chosen to have the same S/N in the two bands.
  The blue dotted line represents the signal measured below $\nu_0$, while the red dashed line is measured above $\nu_0$.
  The black line is the normalized product of the previous two and the vertical grey line highlights the location of its maximum.
}
\end{figure}

Because of the telescope geometry, CHIME's formed beams have strong side lobes in the East-West direction spaced roughly 1 (2) degrees at 800 (400) MHz. 
To unambiguously identify the correct lobe where the source is located, we use the variation of the spacing between lobes as a function of frequency.
This is straightforward for radio sources with a broadband spectrum, which will be detected as broadband only in the correct lobe.
However, it can be challenging for signals with narrow-band spectra that can mimic the telescope response in side lobes, as is often the case with repeating FRBs \citep{ple21}.
Of course, it is theoretically possible to form beams in every potential lobe and select the direction where the signal is strongest.
However, this would be prohibitively computationally expensive to properly sample the rapid spatial variability of the telescope response.
For this reason, we have developed the diagnostic plot visible in Fig.~\ref{fig:sidelobes}, where the frequency bandwidth is divided into two parts so that each one contains half of the signal measured from a certain source. 
A total of 53 adjacent beams are formed in the East-West direction around the best source position spanning 5 degrees on each side, i.e.\ covering a total of 5 telescope lobes at 400\,MHz. 
The signal strength is measured for the two parts of the band independently and the main lobe is identified where the peaks at the two frequencies align. This is highlighted by the normalized product of the signal at the two frequencies.
If the FRB position from the initial 5$\times$5 beam grid is not found to be in the correct lobe, as indicated by the sidelobe diagnostic in Fig.~\ref{fig:sidelobes}, a new grid of 5$\times$5 beams is formed and the localization is calculated again.
Given the low probability of chance alignment due to the small uncertainty regions and the similar DMs, we verify that each FRB source is localized to the correct lobe of the telescope by applying this method to one burst from each FRB source, typically, the one with the broadest or brightest spectrum. 

\subsection{Identification of the host galaxy}
\label{sec:mcmc_z}
The relatively compact design of CHIME, whose maximum baseline is $\sim 100$\,m, and the low observing frequencies ($400$-$800$\,MHz), limit the localization capability of the instrument.
In fact, after accounting for systematic effects, \citet{mic21} estimated a maximum possible localization precision of $\sim 11$\arcsec, insufficient to pinpoint a single host galaxy for most of the detected FRBs.
However, the DM of an FRB can be used to estimate the maximum distance that the source can have assuming models of electron densities in the different media travelled by the radio waves.
FRBs with a low DM excess with respect to Milky Way models will have a low upper limit on their source redshift.
This can be used to place a limit on the redshift of host galaxy candidates in the localization region of an FRB.
To this end, a Bayesian framework was developed by modeling the different contributions to the total FRB DM with priors motivated by previous studies.
The details of this method are provided by \citet{bha21_r4}.
As opposed to their work, we have used as priors on the host DM contribution a log-normal distribution with $\text{mean}=96.4$\,pc\,cm$^{-3}$ and $\text{standard deviation}=0.9$, values based on the work of \citet{2020ApJ...900..170Z}.
The DM distribution of the FRBs detected by CHIME/FRB implies that a small fraction ($\sim 1\%$) will have only one possible host galaxy within his baseband localization region that has a redshift below the maximum limit estimated from its DM.

\subsection{Optical spectroscopic observations}
\label{sec:GTC}
To obtain spectroscopic redshifts for galaxy host candidates, we used the Optical System for Imaging and low-intermediate Resolution Integrated Spectroscopy (OSIRIS)\footnote{\url{http://www.gtc.iac.es/instruments/osiris}} mounted at the Gran Telescopio Canarias (GTC). 
Details about the instrument configuration for this observation and the data analysis are reported in Appendix~\ref{sec:GTC_obs}.

\section{Results}
\label{sec:results}
The positions of the FRB sources measured with baseband data from CHIME/FRB are reported in Table~\ref{tab:loc}.
The position of single events from each repeater can be found in Appendix~\ref{sec:appendix_positions}.
The localization precision and low DM allowed us to obtain an interesting galaxy association for two of the FRB sources reported here, namely FRBs 20180814A and 20190303A.
The localization regions for the rest of the sample contain too many galaxies to draw significant conclusions on plausible host galaxy candidates.

\begin{deluxetable*}{lllllllllll}
\tablecaption{
Sky position of 13 repeating FRB sources discovered by CHIME/FRB localized with their baseband data. 
The source name and other published designations are presented, together with the values of RA and Dec (J2000), their 1$\sigma$ uncertainties, DM, maximum redshift estimated for the host galaxy, and number of bursts with baseband data used in the analysis.
Both RA and Dec uncertainties are in the same units of seconds of arc on the sky, i.e. the uncertainty regions are approximately circular on the sky.
\label{tab:loc}
}
\tablehead{
\colhead{Source} & \colhead{Previous name} & \colhead{RA} & \colhead{$\sigma_\text{RA}$ (\arcsec)} & \colhead{Dec} & \colhead{$\sigma_\text{Dec}$ (\arcsec)} & \colhead{DM (pc\,cm$^{-3}$)} & \colhead{$z_\text{max}$} & \colhead{\#}
}
\startdata
20180814A & 180814.J0422+73\tablenotemark{a} & 4$^h$22$^m$44$^s$ & 18 & 73\degr39\arcmin52\arcsec & 20 & 189.4(4)\tablenotemark{a} & 0.091 & 4 \\
20181128A & 181128.J0456+63\tablenotemark{b} & 4$^h$55$^m$41$^s$ & 47 & 63\degr15\arcmin27\arcsec & 47 & 450.5(3)\tablenotemark{b} & 0.44 & 1 \\
20181119A & 181119.J12+65\tablenotemark{b} & 12$^h$41$^m$52$^s$ & 25 & 65\degr7\arcmin2\arcsec & 29 & 364.05(9)\tablenotemark{b} & 0.43 & 2 \\
20190116A & 190116.J1249+27\tablenotemark{b} & 12$^h$49$^m$8$^s$ & 33 & 27\degr8\arcmin2\arcsec & 33 & 441(2)\tablenotemark{b} & 0.56 & 1 \\
20190222A & 190222.J2052+69\tablenotemark{b} & 20$^h$52$^m$12$^s$ & 15 & 69\degr44\arcmin42\arcsec & 17 & 460.6(2)\tablenotemark{b} & 0.49 & 1 \\
20190208A & 190208.J1855+46\tablenotemark{c} & 18$^h$54$^m$7$^s$ & 12 & 46\degr55\arcmin20\arcsec & 13 & 580.05(15)\tablenotemark{c} & 0.68 & 3 \\
20190604A & 190604.J1435+53\tablenotemark{c} & 14$^h$34$^m$47$^s$ & 28 & 53\degr18\arcmin20\arcsec & 28 & 552.65(5)\tablenotemark{c} & 0.70 & 1 \\
20190213B & 190212.J18+81\tablenotemark{c} & 18$^h$25$^m$2$^s$ & 20 & 81\degr24\arcmin5\arcsec & 26 & 302(1)\tablenotemark{c} & 0.31 & 3 \\
20180908B & 180908.J1232+74\tablenotemark{c} & 12$^h$32$^m$48$^s$ & 38 & 74\degr10\arcmin21\arcsec & 51 & 195.6(2)\tablenotemark{c} & 0.17 & 1 \\
20190117A & 190117.J2207+17\tablenotemark{c} & 22$^h$6$^m$38$^s$ & 13 & 17\degr22\arcmin6\arcsec & 13 & 393.6(8)\tablenotemark{c} & 0.46 & 2 \\
20190303A & 190303.J1353+48\tablenotemark{c} & 13$^h$51$^m$59$^s$ & 11 & 48\degr7\arcmin16\arcsec & 12 & 222.4(7)\tablenotemark{c} & 0.22 & 17 \\
20190417A & 190417.J1939+59\tablenotemark{c} & 19$^h$39$^m$4$^s$ & 15 & 59\degr19\arcmin55\arcsec & 16 & 1378.2(2)\tablenotemark{c} & 1.2 & 3 \\
20190907A & 190907.J08+46\tablenotemark{c} & 8$^h$9$^m$47$^s$ & 40 & 46\degr22\arcmin46\arcsec & 36 & 309.6(2)\tablenotemark{c} & 0.33 & 1 \\
\enddata
\tablenotetext{a}{Presented by \citet{chi19_r2}}
\tablenotetext{b}{Presented by \citet{chi19_8repeaters}}
\tablenotetext{c}{Presented by \citet{fon20}}
\end{deluxetable*}

\subsection{FRB~20180814A}
The low \dm{189.4(4)} of this source together with a significant Galactic disk contribution of
$\text{DM}_\text{MW} = 87$ \citep{cor02_ne2001} to $108$ \citep{yao17_ymw16} pc\,cm$^{-3}$ implies that the host galaxy must be in the local Universe.
Using a marginalized posterior of the host galaxy redshift (see \textsection\ref{sec:mcmc_z}), we obtain a one-sided 95\% Bayesian upper limit on the redshift $z < 0.091$.

\begin{figure}
\epsscale{1.2}
\plotone{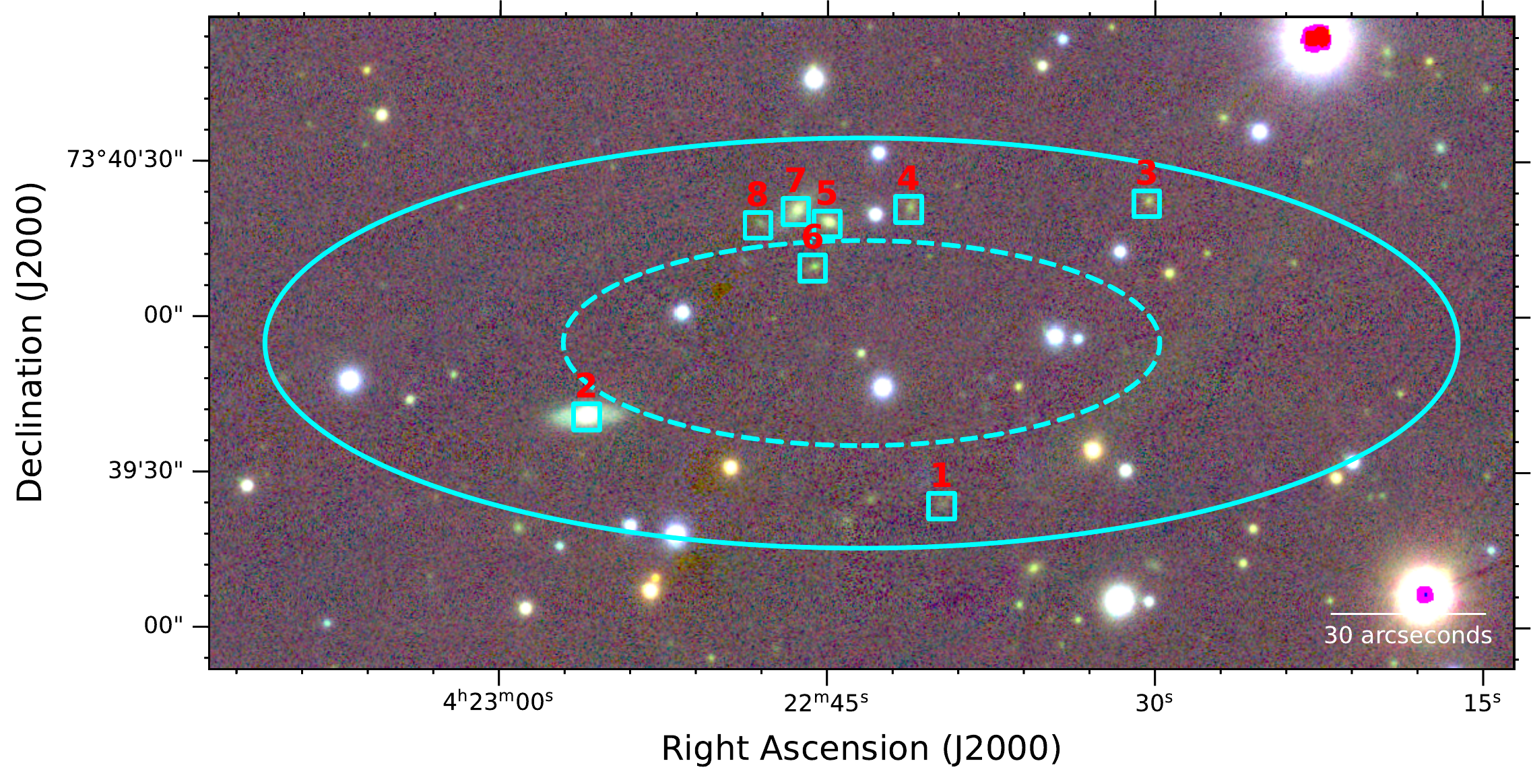}
\caption{PanSTARRS RGB-image of the FRB 20180814A 1- (dotted cyan ellipse) and 2- (solid cyan ellipse) $\sigma$ localization regions. Cyan boxes show the locations of 8 host galaxy candidates within the localization region (see Table~\ref{tab:r2galaxies}) identified in the PanSTARRS data.
We identify source 2 (PanSTARRS-DR1 J042256.01+733940.7) as the most probable host for the FRB.
Pink regions visible in the image have been used to block bright stars.
}
\label{fig:r2field}
\end{figure}

We searched the PanSTARRS-DR1 catalogue \citep{2016arXiv161205560C} for host candidates within the 86\% localization region (corresponding to the $2\sigma$ confidence region for a bivariate normal distribution) of the FRB, and found 8 sources that are listed in Table~\ref{tab:r2galaxies} and plotted in Fig.~\ref{fig:r2field} with the FRB localization region. 

\subsubsection{Spectroscopic redshift}
\begin{figure}
\epsscale{1.2}
\plotone{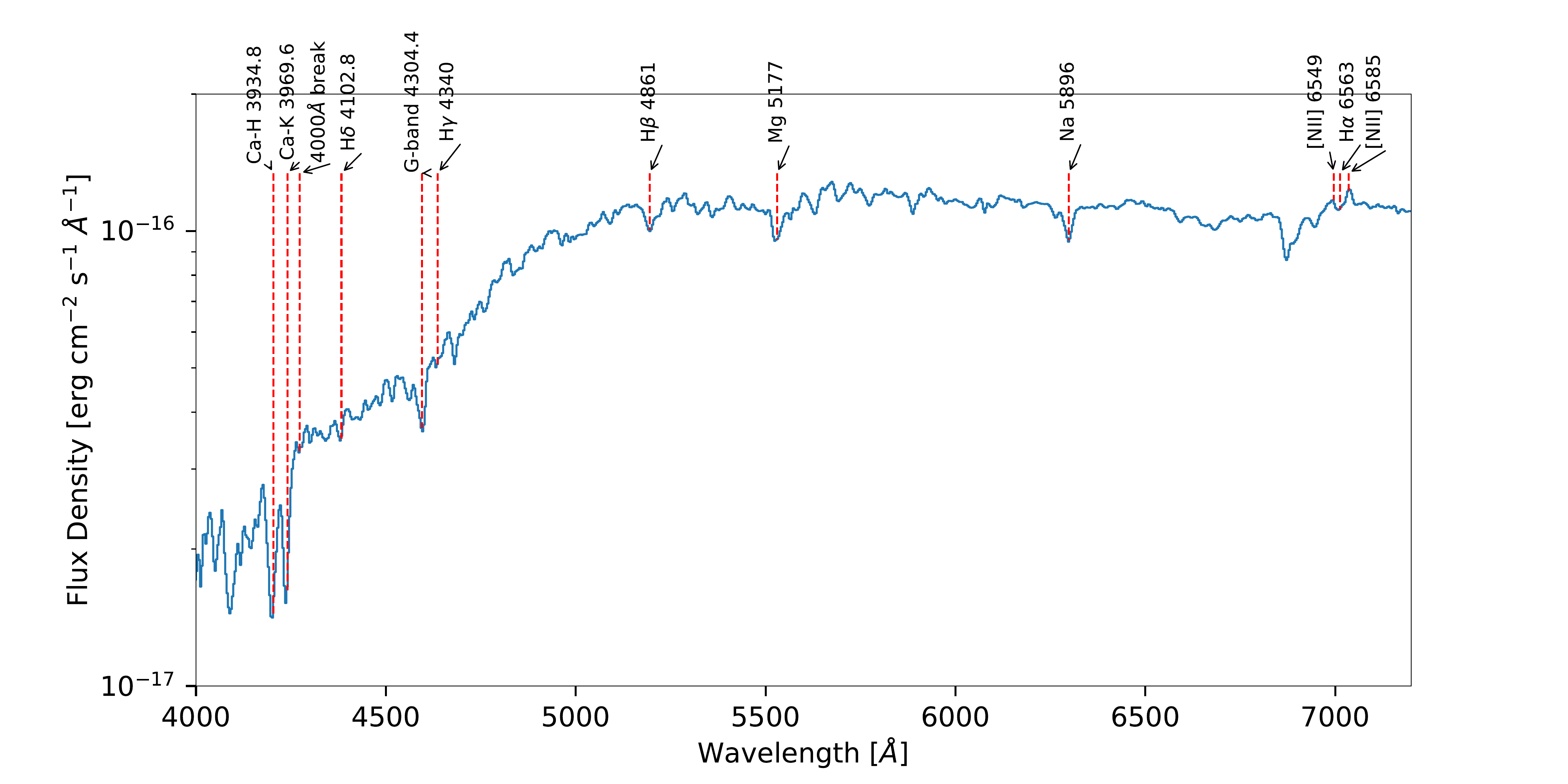}
\caption{
Optical spectrum of PanSTARRS-DR1 J042256.01+733940.7, identified as likely host of FRB 20180814A, measured with GTC/OSIRIS.
}
\label{fig:r2spectrum}
\end{figure}

Most of the galaxies in the field do not have a spectroscopic redshift reported in the literature.
We used the GTC to obtain their redshifts as described in \textsection\ref{sec:GTC} and we report the values we obtained in Table~\ref{tab:r2galaxies}.
We excluded source 8 from our multi-object spectroscopic observations because it is too faint (rK mag = 21.81) to obtain a reliable spectroscopic redshift with our observations. Moreover, its photometric redshift was greater than the maximum value estimated for the FRB host galaxy by 4 standard deviations. 
With a redshift $z=0.06835(1)$, source 2 (PanSTARRS-DR1 J042256.01+733940.7) is the only galaxy in the field satisfying the condition $z < 0.091$.
Therefore, we identify this galaxy as the most probable host for FRB 20180814A.
The spectrum of PanSTARRS-DR1 J042256.01+733940.7 is reported in Fig.~\ref{fig:r2spectrum}.

\begin{deluxetable}{lllll}
\tablecaption{
 Galaxies from the PanSTARRS-DR1 catalogue in the 86\% localization region of FRB 20180814A. Redshifts have been measured using the GTC telescope. Galaxy PanSTARRS-DR1 J042256.01+733940.7, a.k.a.\ source 2, is the most probable host in the field given its low redshift.
 \label{tab:r2galaxies}
}
\tablehead{
 \colhead{\#} & \colhead{RA (J2000)} & \colhead{Dec (J2000)} & \colhead{rK mag\tablenotemark{a}} & \colhead{$z$}
}
\startdata
1 & 4$^h$22$^m$39.77$^s$ & 73\degr39\arcmin23.6\arcsec & 21.28 & 0.412(1) \\
2 & 4$^h$22$^m$56.01$^s$ & 73\degr39\arcmin40.7\arcsec & 17.15 & 0.06835(1)  \\
3 & 4$^h$22$^m$30.38$^s$ & 73\degr40\arcmin22.0\arcsec & 20.75 & 0.408(1) \\
4 & 4$^h$22$^m$41.29$^s$ & 73\degr40\arcmin20.9\arcsec & 20.65 & 0.376(1)  \\
5 & 4$^h$22$^m$45.02$^s$ & 73\degr40\arcmin18.1\arcsec & 19.26 & 0.235(1) \\
6 & 4$^h$22$^m$45.68$^s$ & 73\degr40\arcmin09.5\arcsec & 21.01 & 0.411(1) \\
7 & 4$^h$22$^m$46.46$^s$ & 73\degr40\arcmin20.5\arcsec & 18.75 & 0.237(1) \\
8 & 4$^h$22$^m$48.19$^s$ & 73\degr40\arcmin17.8\arcsec & 21.81 & 0.5(1)\tablenotemark{b}\\
\enddata
\tablenotetext{a}{Kron r-band magnitude.}
\tablenotetext{b}{Photometric redshift from the PanSTARRS catalogue.}
\end{deluxetable}

\subsubsection{Properties of PanSTARRS-DR1 J042256.01+733940.7}
Based on its rest-frame (g-r) color of 0.74 AB mag and absolute r-band magnitude of $-20.75$ AB mag, PanSTARRS-DR1 J042256.01+733940.7 can be classified as a red sequence (early-type) galaxy using the color–magnitude relation identified by \citet{Bell2003ApJS}. 
The WISE color-color classification \citep{wright2010} considers it to be a spiral galaxy with no actively accreting massive black hole in its center, given 
W1 (3.4 $\mu$m) - W2 (4.6 $\mu$m) = $0.13 \pm 0.04$ and W2 (4.6 $\mu$m) - W3 (12 $\mu$m) = $2.45 \pm 0.12$.
The galaxy spectrum reported in Fig.~\ref{fig:r2spectrum} shows multiple Balmer and metal absorption lines, including Calcium H and K lines, the G-band, Mg I, and Na I, while the spectrum lacks prominent emission lines, except for weak [NII] lines, indicating an evolved stellar population lacking young stars and gas. 
This is corroborated by the value of the index $\text{D4000}\approx 1.7$, obtained as the ratio of the flux in the red continuum ($4000-4100$\,\AA) to that in the blue continuum ($3850-3950$\,\AA) in the rest frame, greater than the passive and star-forming galaxy separation cutoff of 1.45 \citep{1999ApJ...527...54B}.
Therefore, PanSTARRS-DR1 J042256.01+733940.7 is likely an early-type spiral galaxy that is currently in its quenched/passive phase and that has stopped forming new stars for more than 1 Gyr \citep{2010MNRAS.405..783M}.

To obtain the physical properties of PanSTARRS-DR1 J042256.01+733940.7, we used a Bayesian inference spectral energy distribution (SED) fitting code, \texttt{Prospector} \citep{Leja2017,prospect2019}.
We fit a delayed-$\tau$ model \citep{simha2014,carnall2019ApJ} whose details are provided in Appendix~\ref{sec:prospector_model}, to 11 broadband optical, near-, and mid-IR filter fluxes. 
The results of this fit are presented in Table~\ref{tab:galactic_properties}.
As visible, the star-formation rate (SFR) is very low, again confirming our classification as a passive galaxy.

\begin{deluxetable*}{llll}
\tablecaption{
 Physical properties of galaxies possibly hosting FRBs identified in this work.
 \label{tab:galactic_properties}
}
\tablehead{
 \colhead{Property} & \colhead{PanSTARRS-DR1} & \colhead{SDSS\tablenotemark{a}} & \colhead{SDSS\tablenotemark{a}} \\
 & \colhead{J042256.01+733940.7} & \colhead{J135159.17+480729.0} & \colhead{J135159.87+480714.2}
}
\startdata
log[SFR] ($\textup{M}_\odot\ \mathrm{yr^{-1}}$) & $< -$0.5 & 0.99(3) &  0.84(4) \\ 
Stellar Metallicity ($\mathrm{log}(\text{Z}/\text{Z}_{\odot})$) & $-$0.61$^{+0.43}_{-0.53}$  & $-$0.39(1) & $-$0.31(7) \\
Stellar mass ($\mathrm{log}(\text{M}/\text{M}_{\odot})$) & 10.78$^{+0.12}_{-0.18}$ & 10.63(3) & 10.75(3) \\
Effective radius (R$_{\mathrm{eff}}$; kpc) & 3.2 & 4.9 & 4.7 \\
Mass-weighted age (Gyr) &  7.6$^{3.3}_{-3.4}$ & 1.72(18) & 4.2(8) \\
(u-r)$_{o}$ (mag) & 2.57$^{+0.17}_{-0.20}$ & 1.79(1) & 1.93(2) \\
A$_{\rm V,young}$ (mag) & 0.43$^{+0.33}_{-0.20}$ & 2.27(13) & 2.44(3) \\
A$_{\rm V,old}$ (mag) & 0.45$^{+0.32}_{-0.18}$ & 0.76(4) & 0.81(1) \\
Absolute r-band mag. (AB) & $-$20.78 & $-$20.49  & $-$19.94 \\
Redshift ($z$) & 0.06835(1) & 0.06386(1) & 0.06437(1) \\
\enddata
\tablenotetext{a}{Properties estimated by the SDSS collaboration \citep{2015ApJS..219...12A}.}
\end{deluxetable*}

\subsection{FRB~20190303A}
The bursts emitted by this FRB source have a relatively small \dm{222.4(7)}.
Through our Bayesian framework (see \textsection\ref{sec:mcmc_z}), we obtained a one-sided 95\% Bayesian upper limit on the redshift $z < 0.22$.

\begin{figure}
\epsscale{1.2}
\plotone{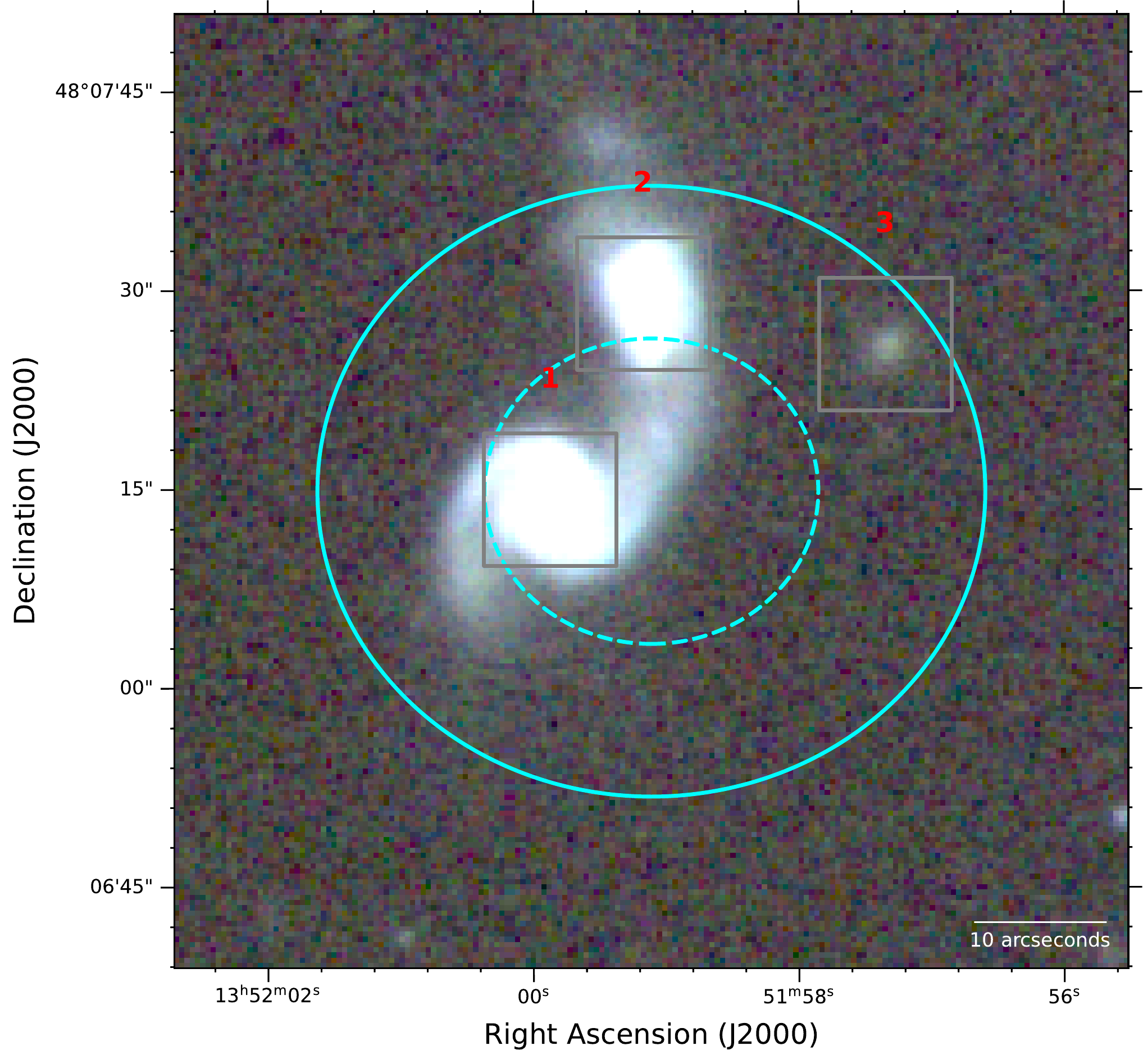}
\caption{SDSS RGB-image of the FRB 20190303A 1- (dotted cyan ellipse) and 2- (solid cyan ellipse) $\sigma$ localization regions. Cyan boxes show the locations of 3 host galaxy candidates within the localization region (see Table~\ref{tab:r17galaxies}) identified in the SDSS data.
}
\label{fig:r17field}
\end{figure}

We queried the Sloan Digital Sky Survey \citep[SDSS;][]{2006AJ....131.2332G} DR12 catalogue \citep{2015ApJS..219...12A} to identify plausible host candidates in the 86\% localization region of the FRB and found three galaxies. These sources are listed in Table \ref{tab:r17galaxies}.
Two of these galaxies are a pair of merging galaxies, SDSS J135159.17+480729.0 and J135159.87+480714.2 (a.k.a.\ MCG+08-25-049 and MCG+08-25-050; \citealp{vor62}), located at a redshift of $z = 0.064$ \citep{ahn12}, that were already noted by \cite{fon20} in the field. 
The SDSS image of the field is shown in Figure \ref{fig:r17field} together with the FRB localization region.

\begin{deluxetable}{lllll}
\tablecaption{
 Galaxies from the DR12 catalogue in the 86\% localization region of FRB 20190303A.
 \label{tab:r17galaxies}
}
\tablehead{
 \colhead{\#} & \colhead{RA (J2000)} & \colhead{Dec (J2000)} & \colhead{r mag\tablenotemark{a}} & \colhead{$z$}
}
\startdata
1 & 13$^h$51$^m$59.$^s$87 & 48\degr7\arcmin14.2\arcsec & 15.50(2) &  0.06386 \\
2 & 13$^h$51$^m$59.$^s$17 & 48\degr7\arcmin29.0\arcsec & 15.99(4) &  0.06437 \\
3 & 13$^h$51$^m$57.$^s$34 & 48\degr7\arcmin25.9\arcsec & 20.12(6) &  0.20(4)\tablenotemark{b}\\
\enddata
\tablenotetext{a}{Petrosian r-band magnitude.}
\tablenotetext{b}{Photometric redshift.}
\end{deluxetable}

We performed a targeted long-slit spectroscopy observation of source 3 with the GTC/OSIRIS (Director’s Discretionary time program GTC04-22ADDT). 
However, the galaxy was too faint to obtain a reliable spectrum.
Source 3 has a photometric redshift measured to be 0.20(4), i.e.\ just compatible with the 95\% maximum limit we estimated for the host galaxy of the FRB.
We calculated the probability of chance association between the FRB and either of the merging galaxies by using a Bayesian framework called Probabilistic Association of Transients to their Hosts \citep[PATH;][]{2021ApJ...911...95A}.
We obtained a posterior probability of true association with either of the merging galaxies $> 0.99$.
Therefore, we consider one of the two merging galaxies in the field to be the likely host of FRB 20190303A.

\subsubsection{Properties of SDSS J135159.17+480729.0 and J135159.87+480714.2}
\begin{figure*}
\plottwo{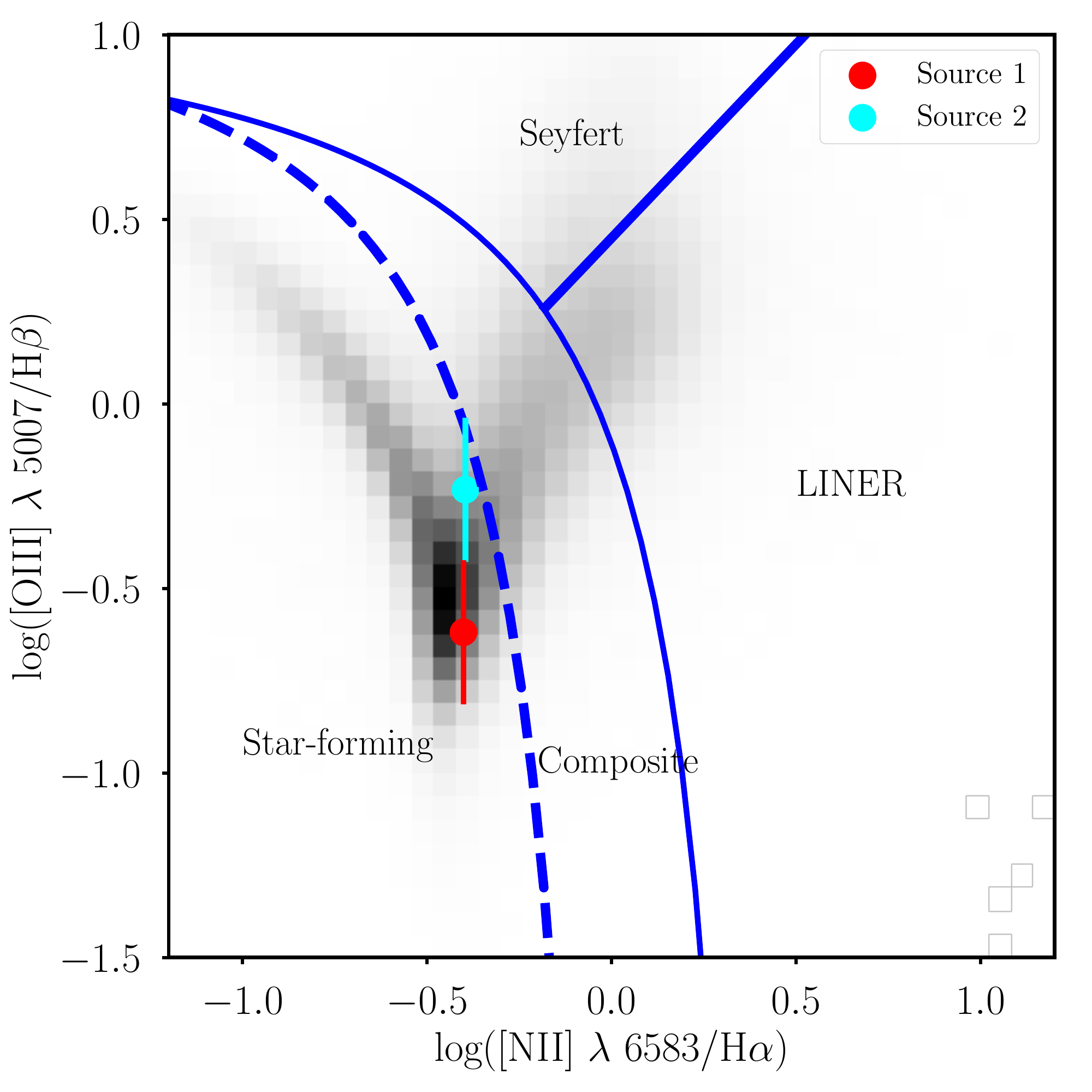}{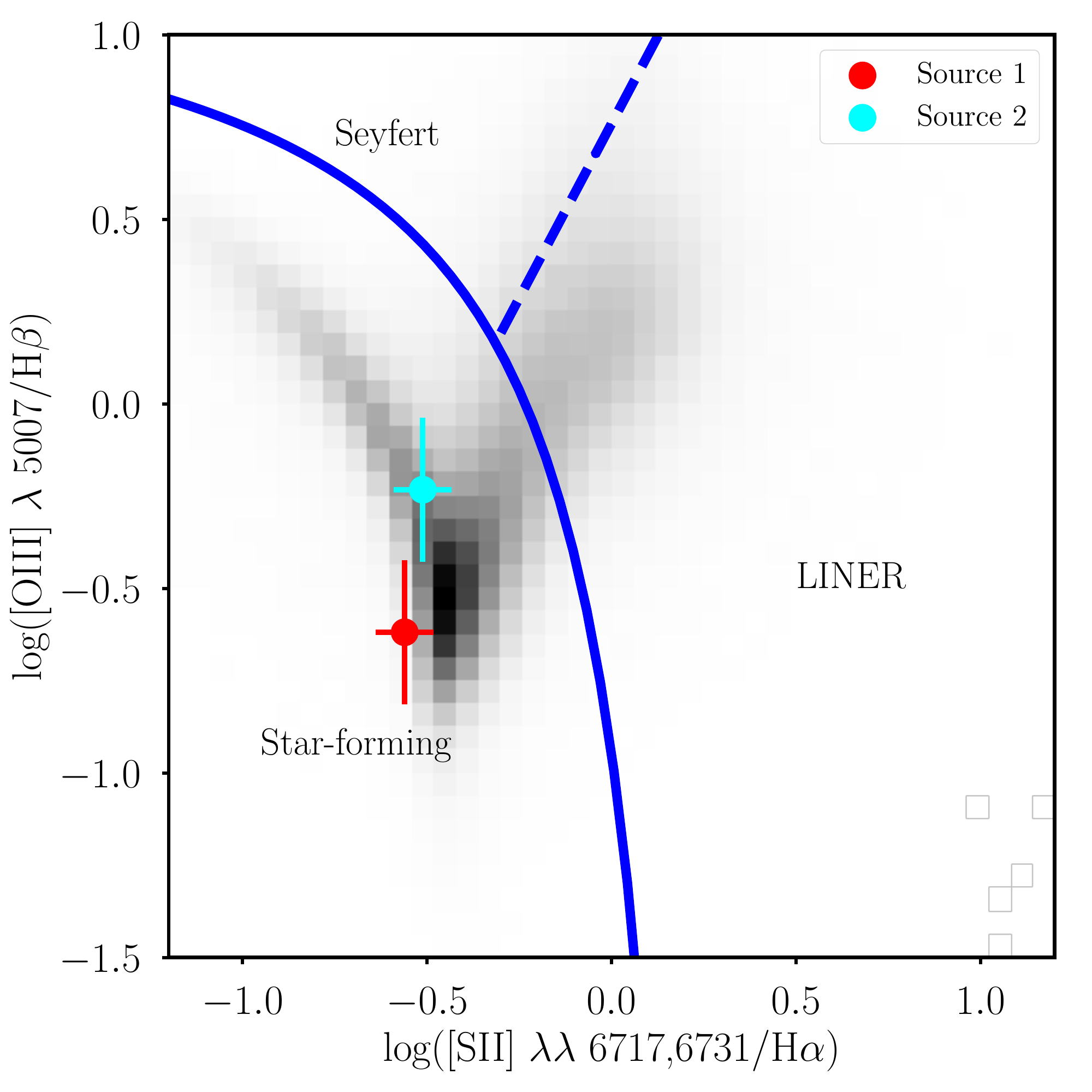}
\caption{BPT diagrams used to classify the merging pair of emission-line galaxies likely hosting FRB~20190303A. 
Left: dashed line shows the \cite{Kauffmann+2003} classification criteria. The \cite{Kewley+2006} classification is shown as the solid line. 
Right:\cite{Kauffmann+2003} criteria is shown as the solid line which separates star-forming galaxies from active galaxies and the dashed lines represent the Seyfert–LINER demarcation from \cite{Schawinski+07}. 
Both SDSS J135159.87+480714.2 (source 1) and SDSS J135159.17+480729.0 (source 2) are classified as star-forming galaxies.
}
\label{fig:bpt_diagrams}
\end{figure*}

The physical properties of the two merging galaxies have been estimated by the SDSS collaboration \citep{2015ApJS..219...12A}; we report the values they obtain in Table~\ref{tab:galactic_properties}.
We used the Baldwin, Philips \& Terlevich (BPT) diagrams of [O III]/H$\beta$ versus [N II]/H$\alpha$ and [O III]/H$\beta$ versus [S II]/H$\alpha$ to classify the dominant source of ionizing radiation in the two galaxies and found that both are classified as star-forming galaxies, as visible in Fig.~\ref{fig:bpt_diagrams}.

\subsection{Other FRB sources}
FRB~20190116A intercepts the line of sight of the Coma cluster. This generated some speculations that the host galaxy might have been part of the cluster \citep{hal19}.
While we are unable to unambiguously identify a host for the FRB, the updated localization allows us to exclude any galaxy in the Coma cluster as a potential host for the FRB after searching multiple optical catalogues \citep{coma_god83,coma_ada06,coma_yag16,coma_mah18}.

FRB~20180908B has a relatively low DM excess, which implies a maximum redshift of $z<0.17$ for its host galaxy in our analysis.
Within the 86\% confidence region of the FRB position, we find 12 galaxies classified in the PanSTARRS DR1 catalogue. 
From their photometric redshifts listed in Table~\ref{tab:r15galaxies}, only a few galaxies are below the maximum limit.
We thus argue that one of them is the host of FRB~20180908B.
However, spectroscopic redshifts are not measured for this galaxy sample and, therefore, the current values could be imprecise.

\begin{deluxetable}{llll}
\tablecaption{
 Galaxies from the PanSTARRS-DR1 catalogue in the 86\% localization region of FRB 20180908B. The photometric redshifts are from the catalogue.
 \label{tab:r15galaxies}
}
\tablehead{
 \colhead{\#} & \colhead{RA (J2000)} & \colhead{Dec (J2000)} & \colhead{$z_\text{photo}$}
}
\startdata
1 & 12$^h$33$^m$32.76$^s$ & 74\degr09\arcmin45.81\arcsec & 0.06(2) \\
2 & 12$^h$33$^m$34.69$^s$ & 74\degr09\arcmin43.22\arcsec & 0.13(1) \\
3 & 12$^h$33$^m$43.43$^s$ & 74\degr10\arcmin05.85\arcsec & 0.15(1) \\
4 & 12$^h$32$^m$09.26$^s$ & 74\degr11\arcmin35.70\arcsec & 0.19(7) \\
5 & 12$^h$33$^m$34.61$^s$ & 74\degr11\arcmin29.55\arcsec & 0.2(5) \\
6 & 12$^h$33$^m$16.95$^s$ & 74\degr09\arcmin13.27\arcsec & 0.20(3) \\
7 & 12$^h$33$^m$09.21$^s$ & 74\degr09\arcmin11.16\arcsec & 0.4(3) \\
8 & 12$^h$32$^m$32.31$^s$ & 74\degr10\arcmin34.71\arcsec & 0.39(10) \\
9 & 12$^h$33$^m$45.03$^s$ & 74\degr10\arcmin11.09\arcsec & 0.4(3) \\
10 & 12$^h$32$^m$40.61$^s$ & 74\degr10\arcmin12.54\arcsec & 0.4(1) \\
11 & 12$^h$32$^m$03.61$^s$ & 74\degr10\arcmin38.23\arcsec & 0.50(5) \\
12 & 12$^h$32$^m$54.57$^s$ & 74\degr09\arcmin16.91\arcsec & 0.59(9) \\
\enddata
\end{deluxetable}

With a \dm{1378.2(2)}, FRB~20190417A has the highest DM of our FRB sample.  
Also, its Faraday rotation has been measured to be relatively high, $\text{RM}\approx4500$\,rad\,m$^{-2}$ \citep{mck22,yi22}.
The high RM could suggest that the source lives in a dense environment, which may be contributing to a large fraction of its total DM \citep{mic18_rm,ann22,dai22}, making the source actually closer than the distance inferred from its DM. 
However, no extended source is detected in the archival PanSTARRS DR1 data, excluding a host galaxy located in the local Universe.
Two FRBs with high RM values have been detected in close proximity of persistent radio sources \citep{cha17_R1,niu22}.
We are currently searching for persistent radio candidates in the field of FRB 20190417A as well as in the field of other FRBs in the sample; the results will be presented in an upcoming paper (Ibik et al., in prep.).

\section{Discussion}
\label{sec:discussion}
The sample of FRB host galaxy candidates exhibits remarkable diversity.
This work shows that even among repeaters, FRBs can reside in environments with very different characteristics.
Together with the other five localized repeaters \citep{hei20}, it is evident that the known repeating FRB hosts exhibit a continuum of properties in terms of their luminosities, stellar masses, metallicity, and star formation rate. 
\citet{Bhandari2022AJ} found tentative evidence for the hosts of repeating FRBs being less massive and less luminous on average, compared to the hosts of apparently non-repeating FRBs. 
The association of the repeating FRBs~20190303A and 20180814A with relatively luminous and massive spiral galaxies seems to contradict this hypothesis.
It is crucial to get more host associations to do meaningful statistical studies of the FRB host population.

\subsection{FRB~20180814A}
The FRB~20180814A host is a spiral galaxy with very low star formation; this is the the first FRB to date associated to a galaxy of this kind.
However, other FRBs have been found to live in environments with old stellar populations. 
Most notably, FRB~20200120E was found to be located in a globular cluster of the M81 galaxy \citep{bha21_m81,kir20}, while FRB~20210117A was recently localized to a dwarf galaxy with little star formation \citep{bha22}. 
These environments suggest a `delayed'-formation channel for some FRB progenitors, such as binary neutron star mergers and accretion-induced white dwarf collapses \citep{2013ApJ...771L..26G,2019MNRAS.484..698R}.
The bursts detected from FRB~20200120E have energies more than three orders of magnitude smaller than those from FRB~20180814A, indicating that `delayed'-formation channels might produce FRB sources with wide ranges of energies.

We re-ran our Bayesian formalism  (\textsection\ref{sec:mcmc_z}) after fixing the redshift of PanSTARRS-DR1 J042256.01+733940.7 and obtained a contribution from the Milky Way disk of \dm[_\text{MW}=]{82(8)}, from its halo of \dm[_\text{halo}=]{40_{-12}^{+11}}, from the IGM of \dm[_\text{IGM}=]{45_{-8}^{+13}}, and from the host galaxy of \dm[_\text{host}=]{19_{-11}^{+17}}. 
However, the line of sight to this galaxy intercepts a few galaxy groups that will also contribute to the number density of free electrons in the IGM \citep{li19}.
This implies that the expected contribution of the host galaxy will be even smaller than the estimated value, as expected for early-type galaxies, where less ionized gas is present \citep[e.g.][]{cha22}.
Thus, FRB~20180814A can be a good candidate to study the circumgalactic medium (CGM) in the foreground galaxy groups along the line of sight \citep{pro19} in future studies.
Additionally, one of FRB~20180814A bursts is found to have a relatively high
$\text{RM}\approx700$\,rad\,m$^{-2}$, where the Milky Way expected contribution is $\text{RM}\approx-45$\,rad\,m$^{-2}$ \citep{mck22}.
The extragalactic RM is likely dominated by the FRB local environment. 
Using the estimated host DM contribution and the observed extragalactic RM, we estimate the average line-of-sight component of the magnetic field to be $\langle\mathrm{B}_{||}\rangle \gtrsim 46$\,$\mu$G.
This value is larger than the magnetic field estimated in the ISM around the Sun \citep[1-5 $\mu$G;][]{wielebinski2005cosmic} and in the Galactic center region \citep[20-40 $\mu$G;][]{wielebinski2005cosmic}. 
Since it is likely dominated by the circumburst medium, its value is probably even higher, as previously observed in other FRBs \citep[e.g.][]{mic18_rm}.
\citet{2018ApJ...861..150P} showed that if the circumburst RM is provided by magnetized stellar winds, we would expect a DM contribution much larger than the observed value (\dm[\gtrsim]{500}). 
On the other hand, if the FRB source is young and the rotation measure is provided by a supernova remnant or post-merger ejecta that is expanding in a constant density host ISM, then the ambient electron number density must be small ($\lesssim 0.1$ pc cm$^{-3}$), compatible with progenitors formed via `delayed'-formation channel models, such as binary neutron star mergers and accretion-induced collapse of white dwarfs \citep{2013ApJ...762L..17P,2016ApJ...830L..38M,2020IJAA...10...28L}.

\subsection{FRB~20190303A}
FRB~20190303A is localized to a pair of merging galaxies.
Other FRBs in the literature have been tentatively associated with merger events.
\citet{Law2020} reported a possible association at 7\% probability \citep{hei20} of FRB~20190614D with a galaxy pair;
\cite{hei20} noted that the FRB~20191001 host is also likely in the process of merging with another galaxy;
\cite{2022ApJ...925L..20K} claimed that the host of FRB~20180916B recently underwent a small merger, resulting in a burst of star formation at the galaxy's outskirts;
finally, \cite{2022arXiv221004680R} reported the discovery of FRB~20220610A in a complex (likely merging) galaxy system. These associations of FRBs with merging systems suggest that merger events could facilitate conditions conducive to the formation of FRB progenitors.
Galaxy mergers often cause enhanced star formation in and at the outskirts of the galaxies, making them a promising site of transients associated with young progenitors formed via prompt channels, like core-collapse supernovae, long gamma ray bursts (LGRBs) and hydrogen-poor superluminous supernovae (SLSNe).

A new execution of our Bayesian framework fixing the redshift of the merging pair resulted in a contribution from the Milky Way disk of \dm[_\text{MW}=]{26(3)}, from its halo of \dm[_\text{halo}=]{42(13)}, from the IGM of \dm[_\text{IGM}=]{49_{-13}^{+34}}, and from the host galaxy of \dm[_\text{host}=]{109_{-36}^{+22}}. 
The relatively large contribution of the host to the DM is likely due to the ongoing merging process of the two galaxies, which can have a significant impact on the CGMs \citep{han18}.
With future observations, the bursts from FRB~20190303A may help us to better understand how the CGM is affected by galaxy mergers.
\citet{mck22} reported that the bursts from FRB~20190303A show a relatively large and rapidly varying RM measured to be between \rotm{-703.40(58) \text{ and } -205.41(42)} \citep{mck22}, with a maximum variation measured to be $|\nabla \rm{RM}|\gtrsim -17$\,rad\,cm$^{-2}$\,day$^{-1}$.
The rapid variability suggests a relatively small size of the Faraday screen and, therefore, it argues for a dominant RM contribution from the environment local to the source as opposed to a larger-scale medium in the merging galaxies.

\subsection{Additional follow-up}
Due to the limited localization precision and the higher DM values, it was not possible to unambiguously identify a host galaxy for the rest of the sources presented in Table~\ref{tab:loc}. We report their best position with the intent for other instruments to study additional bursts eventually emitted by these repeating FRBs. 
The detection of repeating FRBs in nearby galaxies is particularly interesting for follow-up studies at other wavelengths. 
The localization precision reported here is sufficient to observe these sources with most optical and X-ray instruments.
The activity of repeating FRBs is monitored daily by CHIME/FRB; detections are reported and made public within a few hours on a dedicated webpage\footnote{\url{https://www.chime-frb.ca/repeaters}}  and VOEvents are shared within seconds of a detection with any subscriber.\footnote{\url{https://www.chime-frb.ca/voevents}}
We encourage rapid multiwavelength follow-up of the sources presented here, and particularly of active repeaters in nearby galaxies.

\section{Conclusions}
\label{sec:conclusions}
We have presented the best positions of 13 repeating FRB sources currently available with CHIME/FRB. 
We have used channelized voltage data stored at the time of candidate events to map the signal strength on the sky around the source and fit this with a 2D Gaussian approximating the beam response of the telescope \citep{mic21}.
The resulting localization regions have uncertainties of the order of 10\arcsec, not precise enough to unambiguously identify a host galaxy for the majority of FRBs.
However, we use the small excess DM of two FRB sources, 20180814A and 20190303A, to place a limit on the maximum redshift that they can have by using conservative estimates on the Milky Way and extragalactic contributions \citep{bha21_r4}. 
In this way, we are able to identify their likely host galaxies in the localization regions.

FRB~20180814A lives in a quiescent galaxy, while FRB~20190303A is in a merging pair of spiral galaxies undergoing significant star formation.
It has been argued that repeating FRBs may represent a different sub-class of FRBs compared to apparently one-of sources due to their characteristics \citep[e.g.][]{fon20}.
These two diametrically opposed hosts demonstrate that even among repeating FRB sources, there is an incredible variety of different environments that they can inhabit.
These galaxy hosts further reveal the plethora of environments that can harbor FRBs \citep{hei20}.
It is clear that a precise localization of a large number of FRBs is needed to shed a light on the type and number of FRB progenitors.
To this end, CHIME/FRB Outriggers is currently under development to increase the localization capability of CHIME/FRB to $\sim 50$ milliarcseconds \citep{leu21,cas22,men22}.
In the meantime, we encourage follow-ups of repeating FRB sources with other interferometers to increase the localization precision. 
We provide a public web page with updated detection from repeater sources and rapid VOEvents for subscribers.
This is also important for multi-wavelength follow-ups since they often require a precise position and are particularly constraining when observing nearby repeating FRBs, as it is the case for FRBs 20180814A and 20190303A.

%

\vspace{5mm}
\facilities{GTC, CHIME}


\software{
astropy \citep{2018AJ....156..123A}, 
bitshuffle \citep{2017ascl.soft12004M},
cython \citep{behnel2011cython},
emcee \citep{2013PASP..125..306F},
GTCMOS \citep{gtcmos},
hdf5 \citep{hdf5}, 
IRAF \citep{tody1986iraf,tody1993iraf},
matplotlib \citep{2007CSE.....9...90H}, 
numpy \citep{2020Natur.585..357H},
PATH \citep{2021ApJ...911...95A},
prospector \citep{Leja2017,prospect2019},
scipy \citep{2020NatMe..17..261V}
}

\acknowledgments
We acknowledge that CHIME is located on the traditional, ancestral, and unceded territory of the Syilx/Okanagan people. We are grateful to the staff of the Dominion Radio Astrophysical Observatory, which is operated by the National Research Council of Canada.  CHIME is funded by a grant from the Canada Foundation for Innovation (CFI) 2012 Leading Edge Fund (Project 31170) and by contributions from the provinces of British Columbia, Qu\'{e}bec and Ontario. The CHIME/FRB Project is funded by a grant from the CFI 2015 Innovation Fund (Project 33213) and by contributions from the provinces of British Columbia and Qu\'{e}bec, and by the Dunlap Institute for Astronomy and Astrophysics at the University of Toronto. Additional support was provided by the Canadian Institute for Advanced Research (CIFAR), McGill University and the McGill Space Institute thanks to the Trottier Family Foundation, and the University of British Columbia. The Dunlap Institute is funded through an endowment established by the David Dunlap family and the University of Toronto. 
FRB research at UBC is funded by an NSERC Discovery Grant and by the Canadian Institute for Advanced Research.  
The CHIME/FRB baseband system is funded in part by a Canada Foundation for Innovation John R. Evans Leaders Fund award to IHS. 
This work is also based on observations made with the Gran Telescopio Canarias (GTC), installed at the Spanish Observatorio del Roque de los Muchachos of the Instituto de Astrofísica de Canarias, in the island of La Palma.
A.B.P. is a McGill Space Institute (MSI) Fellow and a Fonds de Recherche du Quebec - Nature et Technologies (FRQNT) postdoctoral fellow.
A.M.C is supported by a NSERC Doctoral Postrgradute Scholarship.
B.M.G. acknowledges the support of the Natural Sciences and Engineering Research Council of Canada (NSERC) through grant RGPIN-2022-03163, and of the Canada Research Chairs program.
C.L. was supported by the U.S. Department of Defense (DoD) through the National Defense Science \& Engineering Graduate Fellowship (NDSEG) Program.
F.A.D is supported by the U.B.C Four Year Fellowship.
K.S. is supported by the NSF Graduate Research Fellowship Program.
K.W.M. holds the Adam J. Burgasser Chair in Astrophysics and is supported by an NSF Grant (2008031).
M.B. is a McWilliams Fellow.
M.D. is supported by a Canada Research Chair, NSERC Discovery Grant, CIFAR, and by the FRQNT Centre de Recherche en Astrophysique du Qu\'ebec (CRAQ).
P.S. is a Dunlap Fellow.
S.P.T. is a CIFAR Azrieli Global Scholar in the Gravity and Extreme Universe Program.
V.M.K. holds the Lorne Trottier Chair in Astrophysics \& Cosmology, a Distinguished James McGill Professorship, and receives support from an NSERC Discovery grant (RGPIN 228738-13), from an R. Howard Webster Foundation Fellowship from CIFAR, and from the FRQNT CRAQ.
Z.P. is a Dunlap Fellow.



\appendix

\section{GTC/OSIRIS observations of host galaxy candidates}
\label{sec:GTC_obs}
In order to measure the spectroscopic redshift of the 7 host galaxy candidates in the FRB 20180814A 2-$\sigma$ localization region (see Fig.\ref{fig:r2field}), we performed an observation with the the GTC/OSIRIS.
The summary of the observation is given in Table~\ref{tab:gct_log}.
To obtain the spectra of these targets simultaneously, we utilized the multi-object spectroscopy (MOS) mode. 
The mask was designed with the OSIRIS Mask Designer Tool \citep{mask1,mask2}, using a set of five fiducial stars and catalogue coordinates of the host galaxy candidates. The observations were performed with the R500B grism covering the spectral range 3600$-$7200 {\AA}. For the target galaxies we used rectangular slitlets with length between 1.5\arcsec\ and 20\arcsec\ and a width of 1.5\arcsec.
The spectral resolution for this observation was $\sim$21 \AA. 

The obtained spectra were reduced using the GTCMOS pipeline \citep{gtcmos} and standard IRAF routines \citep{tody1986iraf,tody1993iraf}. All spectra were bias-subtracted and flat-fielded. For flux calibration we used the spectrophotometric standard G191-B2B \citep{std2,standard,massey1988} observed during the same night as the targets. 
A set of arc-lamp spectra of Ne, Hg and Ar was used for wavelength calibration. 
The rms errors of the resulting solutions were $<$2 Å. 

The resulting product of the reduction contained 2D calibrated
spectra collected in all of the slitlets. We extracted each spectrum, identified lines for every galaxy and estimated their redshifts.
We then verified our results by comparing the extracted spectra with the galaxy templates from the Manual and Automatic Redshifting
Software \citep[MARZ;][]{Marz}. The corresponding redshifts are presented in Table \ref{tab:r2galaxies}.

\begin{deluxetable*}{lllllllll}
\tablecaption{
    Log of the GTC/OSIRIS spectroscopic observation.
    \label{tab:gct_log}
}
\tablehead{
  \colhead{Program} & \colhead{Date} & \colhead{Mode} & \colhead{Grism} & \colhead{Position} & \colhead{Exposure} & \colhead{Seeing} & \colhead{Airmass} & \colhead{Night}
}
\startdata
 GTCMULTIPLE3B-20BMEX & 12/02/2021 & MOS & R500B & 0\degr & 3 $\times$ 1200 s & 0.7$\arcsec$  & 1.47-1.52  & Dark   \\
\enddata
\end{deluxetable*}

\section{Localization of single bursts}
\label{sec:appendix_positions}
Sky positions of repeating FRBs presented in Table~\ref{tab:loc} have been obtained as weighted averages of single bursts detected for each source.
In Table~\ref{tab:loc_single}, we report the localization of single bursts corrected for the systematic effects described by Eq.~\ref{eq:loc_calibration}.
One burst from FRB 20180814A, 20190611A, detected on MJD 58645.78660, was not included due to processing issues.

\startlongtable
\begin{deluxetable}{llllll}
\tablecaption{
Localization of single bursts (J2000) from repeating FRB sources reported in Table~\ref{tab:loc} with the same units.
Modified Julian dates (MJDs) represent UTC topocentric arrival times at CHIME referenced to 400\,MHz using $k_{\text{DM}}^{-1} = 2.41 \times 10^{-4}$\,cm$^{-3}$\,pc\,MHz$^{-2}$\,s$^{-1}$ and the source DMs reported in Table~\ref{tab:loc_single}. The uncertainty on MJD values is $\lesssim 0.2$\,s.
\label{tab:loc_single}}
\tabletypesize{\footnotesize}
\tablehead{
\colhead{FRB} & \colhead{MJD} & \colhead{RA} & \colhead{$\sigma_\text{RA}$} & \colhead{Dec} & \colhead{$\sigma_\text{Dec}$} \\
\colhead{} & \colhead{} & \colhead{(J2000)} & \colhead{(\arcsec)} & \colhead{(J2000)} & \colhead{(\arcsec)} \\
}
\startdata
\multicolumn{6}{c}{FRB~20180814A} \\
20190625E & 58659.738741 & 4$^h$22$^m$40$^s$ & 54 & 73\degr41\arcmin10\arcsec & 49 \\
20190626A & 58660.776313 & 4$^h$22$^m$40$^s$ & 47 & 73\degr40\arcmin11\arcsec & 49 \\
20191029A & 58785.404155 & 4$^h$22$^m$48$^s$ & 20 & 73\degr39\arcmin21\arcsec & 25 \\
20191111A & 58798.371711 & 4$^h$22$^m$35$^s$ & 32 & 73\degr40\arcmin27\arcsec & 40 \\
\hline
\multicolumn{6}{c}{FRB~20181128A} \\
20201215C & 59198.302165 & 4$^h$55$^m$41$^s$ & 47 & 63\degr15\arcmin27\arcsec & 47 \\
\hline
\multicolumn{6}{c}{FRB~20181119A} \\
20200621C & 59021.110446 & 12$^h$41$^m$52$^s$ & 27 & 65\degr7\arcmin9\arcsec & 31 \\
20201204D & 59187.659263 & 12$^h$41$^m$53$^s$ & 55 & 65\degr6\arcmin22\arcsec & 68 \\
\hline
\multicolumn{6}{c}{FRB~20190116A} \\
20190116A & 58499.546925 & 12$^h$49$^m$8$^s$ & 33 & 27\degr8\arcmin2\arcsec & 33 \\
\hline
\multicolumn{6}{c}{FRB~20190222A} \\
20190301A & 58543.752114 & 20$^h$52$^m$12$^s$ & 15 & 69\degr44\arcmin42\arcsec & 17 \\
\hline
\multicolumn{6}{c}{FRB~20190208A} \\
20200124A & 58872.779844 & 18$^h$54$^m$8$^s$ & 25 & 46\degr55\arcmin58\arcsec & 28 \\
20200513B & 58982.475387 & 18$^h$54$^m$10$^s$ & 29 & 46\degr55\arcmin3\arcsec & 31 \\
20210203B & 59248.748741 & 18$^h$54$^m$7$^s$ & 13 & 46\degr55\arcmin19\arcsec & 13 \\
\hline
\multicolumn{6}{c}{FRB~20190604A} \\
20190606A & 58640.232311 & 14$^h$34$^m$47$^s$ & 28 & 53\degr18\arcmin20\arcsec & 28 \\
\hline
\multicolumn{6}{c}{FRB~20190213B} \\
20191217A & 58834.360372 & 18$^h$25$^m$14$^s$ & 26 & 81\degr23\arcmin9\arcsec & 45 \\
20200725B & 59055.235372 & 18$^h$24$^m$57$^s$ & 31 & 81\degr24\arcmin41\arcsec & 46 \\
20210216B & 59261.715524 & 18$^h$24$^m$46$^s$ & 35 & 81\degr24\arcmin20\arcsec & 39 \\
\hline
\multicolumn{6}{c}{FRB~20180908B} \\
20190621A & 58655.098197 & 12$^h$32$^m$48$^s$ & 38 & 74\degr10\arcmin21\arcsec & 51 \\
\hline
\multicolumn{6}{c}{FRB~20190117A} \\
20190117A & 58500.929541 & 22$^h$6$^m$38$^s$ & 13 & 17\degr22\arcmin5\arcsec & 14 \\
20191223A & 58840.004946 & 22$^h$6$^m$38$^s$ & 28 & 17\degr22\arcmin19\arcsec & 36 \\
\hline
\multicolumn{6}{c}{FRB~20190303A} \\
20190702B & 58666.135223 & 13$^h$52$^m$0$^s$ & 13 & 48\degr7\arcmin26\arcsec & 13 \\
20191013A & 58769.855170 & 13$^h$52$^m$0$^s$ & 46 & 48\degr6\arcmin39\arcsec & 48 \\
20191020A & 58776.825593 & 13$^h$52$^m$6$^s$ & 25 & 48\degr7\arcmin10\arcsec & 26 \\
20191110A & 58797.772937 & 13$^h$51$^m$59$^s$ & 16 & 48\degr7\arcmin32\arcsec & 17 \\
20191113A & 58800.758656 & 13$^h$51$^m$58$^s$ & 19 & 48\degr7\arcmin2\arcsec & 20 \\
20191116A & 58803.756845 & 13$^h$51$^m$58$^s$ & 12 & 48\degr7\arcmin20\arcsec & 12 \\
20191117A & 58804.762867 & 13$^h$52$^m$4$^s$ & 49 & 48\degr7\arcmin45\arcsec & 61 \\
20191215A & 58832.676200 & 13$^h$52$^m$2$^s$ & 22 & 48\degr6\arcmin56\arcsec & 25 \\
20191231A & 58848.637383 & 13$^h$51$^m$57$^s$ & 27 & 48\degr7\arcmin22\arcsec & 31 \\
20200112A & 58860.601787 & 13$^h$51$^m$60$^s$ & 38 & 48\degr6\arcmin10\arcsec & 41 \\
20200622A & 59022.161027 & 13$^h$51$^m$59$^s$ & 31 & 48\degr7\arcmin43\arcsec & 33 \\
20200809G & 59070.028066 & 13$^h$52$^m$3$^s$ & 14 & 48\degr7\arcmin10\arcsec & 14 \\
20200909A & 59101.937544 & 13$^h$51$^m$57$^s$ & 34 & 48\degr6\arcmin37\arcsec & 30 \\
20210203C & 59248.535427 & 13$^h$51$^m$56$^s$ & 21 & 48\degr7\arcmin6\arcsec & 22 \\
20210207A & 59252.526658 & 13$^h$51$^m$56$^s$ & 23 & 48\degr7\arcmin48\arcsec & 23 \\
20210209B & 59254.523228 & 13$^h$51$^m$58$^s$ & 13 & 48\degr7\arcmin2\arcsec & 13 \\
20210302C & 59275.466332 & 13$^h$51$^m$56$^s$ & 35 & 48\degr7\arcmin48\arcsec & 39 \\
\hline
\multicolumn{6}{c}{FRB~20190417A} \\
20190806A & 58701.275809 & 19$^h$39$^m$12$^s$ & 100 & 59\degr18\arcmin46\arcsec & 78 \\
20200726D & 59056.306742 & 19$^h$39$^m$6$^s$ & 16 & 59\degr19\arcmin58\arcsec & 17 \\
20210304B & 59277.696671 & 19$^h$38$^m$57$^s$ & 24 & 59\degr19\arcmin51\arcsec & 29 \\
\hline
\multicolumn{6}{c}{FRB~20190907A} \\
20200729A & 59059.820902 & 8$^h$9$^m$47$^s$ & 40 & 46\degr22\arcmin46\arcsec & 36 \\
\hline
\enddata
\end{deluxetable}

\section{\texttt{Prospector} fitting model}
\label{sec:prospector_model}
\begin{figure}
\epsscale{1.2}
\plotone{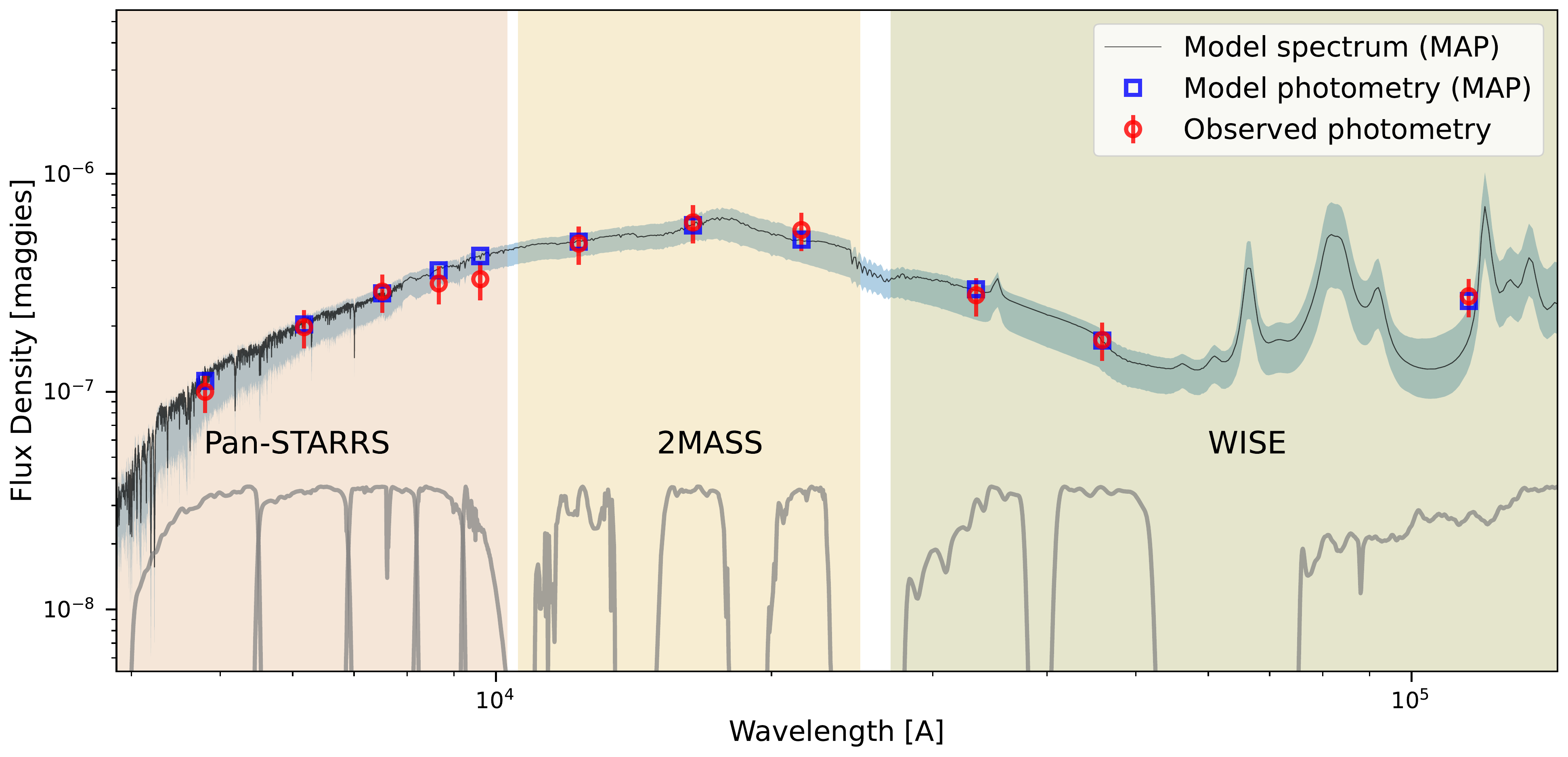}
\caption{\label{fig:prospector}
  Spectral energy distribution of PanSTARRS-DR1 J042256.01+733940.7, identified as likely host of FRB 20180814A. The flux densities in different optical and infrared bands are plotted along with the best-fit {\tt Prospector} model spectrum. The modelled (blue square) and observed (red circle) photometry data are also shown.
}
\end{figure}

We used a python-based Bayesian inference code, {\tt Prospector} \citep{Leja2017,prospect2019}, to estimate major physical properties of galaxy PanSTARRS-DR1 J042256.01+733940.7, identified as likely host of FRB 20180814A.
{\tt Prospector} computes galaxy attributes using stellar population synthesis models provided in the Flexible Stellar Populations Synthesis (FSPS) stellar population code \citep{Conroy2009}. 
We used the MCMC framework (via \texttt{emcee}) of {\tt Prospector} to fit the observed spectral energy distributions (SEDs) and to compute posterior distribution for all free-parameters. 
We used 11 broadband filters listed in Table \ref{tab:prospector_filters} to estimate stellar mass, star-formation rate (95\% confidence upper limit), stellar metallicity, and mass-weighted stellar population age of the galaxy, togheter with dust attenuation due to birth cloud and diffuse dust screens. 
The best-fit SED profile is shown in Fig.~\ref{fig:prospector}.

\begin{deluxetable}{llll}
\tablecaption{
 Broadband filters used to model the SED of galaxy PanSTARRS-DR1 J042256.01+733940.7, possible host of FRB 20180814A.
 \label{tab:prospector_filters}
}
\tablehead{
  \colhead{Instrument} & \colhead{Filter} & \colhead{Wavelength (\AA)} & \colhead{Flux density\tablenotemark{a}}
}
\startdata
PanSTARRS & g & 4810 & 9.98 $\times 10^{-8}$ \\
 & r & 6170 & 1.98 $\times 10^{-7}$  \\
 & i & 7520 & 2.88 $\times 10^{-7}$  \\
 & z & 8660 & 3.14 $\times 10^{-7}$ \\
 & y & 9620 & 3.28 $\times 10^{-7}$  \\
2MASS & J & 12319 & 4.79 $\times 10^{-7}$   \\
 & H & 16420 & 5.98 $\times 10^{-7}$   \\
 & Ks & 21567 & 5.52 $\times 10^{-7}$   \\
WISE & W1 & 33461 & 2.77 $\times 10^{-7}$   \\
 & W2 & 45952 & 1.73 $\times 10^{-7}$  \\
 & W3 & 115526 & 2.74 $\times 10^{-7}$
\enddata
\tablenotetext{a}{
Flux densities are expressed in maggies, where $1\, \text{maggie} = 1\,\text{Jansky} / 3631$. Flux densities at $\lambda < 100000$\,\AA\ are corrected for Galactic extinction according to the prescription of \citet{schlafly2011ApJ}. All flux densities are assigned a 20\% fractional uncertainty, which is larger than the catalogued error.
}
\end{deluxetable}

All flux densities are estimated after correcting for the Milky Way extinction. We fit a delayed-$\tau$ star formation history model \citep{simha2014,carnall2019ApJ} that has nine free parameters described in Table~\ref{tab:prospector_parameters}.
In this model, the star-formation history is proportional to $t\times \exp(-t /\tau)$, where $t$ is the time since the formation epoch of the galaxy, and $\tau$ is the characteristic decay time of our star-formation history. Additionally, we enabled the dust emission model by \cite{draine2007} in the FSPS framework which has three free parameters which regulate the shape of the IR SED: duste(U$_{\rm min}$), duste(Q$_{\rm PAH}$), and  duste($\gamma$). Specifically, duste(U$_{\rm min}$) represents the minimum starlight intensity to which the dust mass is exposed, duste($\gamma$) represents the fraction of dust mass which is exposed to this minimum starlight intensity, and duste(Q$_{\rm PAH}$) quantifies the fraction of total dust mass that is in polycyclic aromatic hydrocarbons (PAHs). 
To account for dust attenuation, we use the two-component \citep{2000ApJ...539..718C} dust attenuation model, which postulates separate birth-cloud (dust1) and
diffuse dust (dust2) screens. In order to estimate the `dust1' parameter, we used an in-built prospector function ${\rm \tt models.transforms.dustratio\_to\_dust1}$.
All parameters are given standard {\tt Prospector} priors (see Table \ref{tab:prospector_parameters}).
To estimate the best-fitted mass-weighted stellar population age value, we used an in-built prospector function { \rm \tt parametric\_mwa}.
We estimated separately internal dust extinction due to young stars (A$_{\rm V,young}$) and old stars (A$_{\rm V,old}$). Finally, we used the SED templates produced via MCMC simulation and estimate the rest frame u-r colour. 
The major physical properties of the galaxy that we obtained are provided in Table~\ref{tab:galactic_properties}. 

\begin{deluxetable*}{lll}
\tablecaption{
 Free parameters fitted with a `delayed-$\tau$' model with {\tt Prospector}.
 \label{tab:prospector_parameters}
}
\tablehead{
  \colhead{Parameter} & \colhead{Description} & \colhead{Prior}
}
\startdata
log(M/M$_{\odot}$)   &  Total stellar mass formed & log-uniform: min=8, max=12 \\
 log(Z/Z$_{\odot}$)  &   Stellar metallicity   & top-hat: min=-2, max=0.2 \\
 dust2 & Diffuse V-band dust optical depth  &  top-hat: min=0, max=3\\
 dust-ratio & Ratio of additional optical depth in the direction & clipped-normal: \\
 & of young stars to diffuse optical depth in all stars & mean=1, sigma=0.3, min=0, max=2 \\
 t$_{\rm age} \rm{~[Gyrs]}$   & Stellar population age of Source 2 & top-hat: min=0.001, max=13.6\\
 $\tau \rm{~[Gyrs]}$ & E-folding time of the SFH   & log-uniform: min=0.1, max=30\\
 duste(U$_{\rm min}$) & From \citep{draine2007} dust attenuation model & top-hat: min=0.1, max=25\\
 duste(Q$_{\rm PAH}$) & From \citep{draine2007} dust attenuation model & top-hat: min=0.5, max=7\\
 duste($\gamma$) & From \citep{draine2007} dust attenuation model & log-uniform: min=0.001, max=0.15 \\
\enddata
\end{deluxetable*}


\bibliography{sample63}{}
\bibliographystyle{aasjournal}



\end{document}